\documentclass{Interspeech2024}
\usepackage{multirow}

\interspeechcameraready 

\title{On the Use of Audio to Improve Dialogue Policies}

\name[affiliation={1}]{Daniel}{Roncel Díaz}
\name[affiliation={1}]{Federico}{Costa}
\name[affiliation={2,1}]{Javier}{Hernando}

\address{
  $^1$Universitat Politècnica de Catalunya, Spain\\
  $^2$Barcelona Supercomputing Center, Spain}
\email{daniel.roncel@estudiantat.upc.edu, federico.costa@upc.edu, javier.hernando@bsc.es}


\begin{document}

\maketitle

\begin{abstract}
    
    With the significant progress of speech technologies, spoken goal-oriented dialogue systems are becoming increasingly popular. One of the main modules of a dialogue system is typically the dialogue policy, which is responsible for determining system actions. This component usually relies only on audio transcriptions, being strongly dependent on their quality and ignoring very important extralinguistic information embedded in the user’s speech. In this paper, we propose new architectures to add audio information by combining speech and text embeddings using a Double Multi-Head Attention component. Our experiments show that audio embedding-aware dialogue policies outperform text-based ones, particularly in noisy transcription scenarios, and that how text and audio embeddings are combined is crucial to improve performance. We obtained a $9.8\%$ relative improvement in the User Request Score compared to an only-text-based dialogue system on the DSTC2 dataset \footnote{The code used to obtain the results, the URLs of the pre-trained models, and the instructions for reproducibility can be found in this repository: \url{https://github.com/danielroncel/tfm}}.
    
\end{abstract}

\section{Introduction}

A dialogue system is a computer program that communicates with a user in a natural way \cite{DSA-2013}. 
Typically, the communication between the user and the system is either text-based or speech-based. 
However, recent multimodal dialogue systems can use both, or also process input images and even body gestures \cite{VID-2017}.
According to their purpose, dialogue systems can be divided into three groups: goal-oriented, question-answering and social dialogue systems \cite{NAT-2018}.
Due to the new advances in deep learning, particularly about contextualizing AI and prompting Large Language Models (LLMs), it has become relatively straightforward to build simple question-answering and social dialogue systems \cite{CBG-2024}.

Goal-oriented spoken dialogue systems are designed to assist the user in solving a particular task, where the user interacts with the application using their voice, typically in a turn-taking conversation. 
This approach has transformed how users manage their daily tasks, offering a more natural and efficient experience.
These systems typically consist of various modules. 
One of the more important modules is the dialogue policy.
Given the transcription of the user's message (or variables inferred from their transcriptions) and the system's latest messages, the dialogue policy determines what the next system action should be. 

In recent years, neural networks have been increasingly utilized to design dialogue policies \cite{ASO-2017, RAI-2023}.
Convolutional Neural Networks (CNNs) were used to extract text features in several works \cite{OTO-2019, SCF-2020}. 
In \cite{ANB-2016}, a Recurrent Neural Network (RNN) is employed, performing a given task very competitively across several metrics when trained on only a few hundred dialogues.
In \cite{ETE-2020}, a more complex system is presented, with GPT-2 \cite{LMA-2019} as its core. 
This large language model uses the dialogue history, dialogue state history, system actions, and response vectors as inputs.
Among the various outputs of this fine-tuned GPT-2 model, it also predicts the next system dialogue act.
To extract text representations, word embeddings have become a common strategy in many Natural Language Processing (NLP) tasks.
Word2Vec \cite{EEO-2013} and GloVe \cite{GGV-2014} are some of the most widely adopted.
Nevertheless, these techniques have certain limitations when applying them to different tasks \cite{HCA-2019}.
Recent works have created context-sensitive word representations, trying to capture highly transferable and task-agnostic properties of language \cite{LKA-2019}. 
ELMo \cite{DCW-2018}, BERT \cite{BPT-2018} and GPT-2 \cite{LMA-2019} are self-supervised deep neural language models that can be fine-tuned to create models for a wide range of downstream NLP tasks.

The Automatic Speech Recognition (ASR) module is crucial to building a successful dialogue policy.
Nevertheless, using only text features does not guarantee extracting all the information conveyed by the user while speaking.
To mitigate the impact of poor transcriptions and to extract information exclusively found in the original audio, researchers have experimented with audio-aware dialogue policies \cite{AEH-2021, AEA-2022, DSW-2020}.
In \cite{AEA-2022}, they proposed a novel dialogue policy implementation which uses the transcript of the last conversation turns as input but also the last audio of the user.
In their experiments, they provide evidence that making the dialogue policy audio-aware improves the performance of the system, especially in scenarios in which the outputs of the ASR are noisy.

The choice of how to combine multimodal features is crucial for maximizing the performance of deep learning models \cite{ETF-2023}.
In \cite{DMH-2021}, the authors developed a Double Multi-Head Attention System, achieving state-of-the-art results for Multimodal Speech and Text Emotion Recognition \cite{DMH-2024, BAE-2024}.
In this paper, we propose an audio-aware dialogue policy enhanced with an early fusion strategy using the Double Multi-Head Attention.
First, a Multi-Head Attention (MHA) layer transforms multimodal representations into complementary contextualized representations. 
Then, a second attention mechanism is applied to pool these representations into a single vector.
Text representations are generated using GPT-2 \cite{LMA-2019} and speech representations are extracted from the raw audio waveforms using Wav2Vec2.0 \cite{WAF-2020}, HuBERT \cite{HSS-2021}, UniSpeechSAT \cite{UUS-2022} or WavLM \cite{WLS-2022} pre-trained self-supervised models.
We consider architectures developed in \cite{AEA-2022} as baselines and we show that our strategy improves the system's performance considerably.

The rest of this paper is structured as follows. 
Section 2 details our proposed system.
In Section 3 we describe our experiments, where experimental setups and results are included. 
Concluding remarks are given in Section 4.

\section{System Description}

\subsection{Dialogues pre-processing}

Generally, task-oriented dialogue systems are based on slot-filling and dialogue acts. 
Slot-filling means that before solving the task, the system must collect some information from the user to later provide a solution that satisfies the user's requirements. 
Dialogue acts refers to the actions that either the user or the system are performing in each interaction.
Each dialogue act is formed by an action and, if the action is related to a slot, the name of the slot itself.
In a single turn, the dialogue policy can output one or more dialogue acts.
The dialogue acts of the system's last turn is considered as the label.
If the system outputs several dialogue acts, the label is the concatenation of them.

The dialogue policies that we trained use the transcription of the last nine turns of the conversation as textual input.
Three sequences of embeddings are constructed: text dialogue history embeddings, source (user/system) embeddings and position embeddings.
Once constructed, the three sequences of embeddings are summed and the resulting sequence is passed through the GPT-2 model.

To build the text dialogue history embeddings, $<$user$>$ and $<$system$>$ tokens are added at the beginning of each user and system message, respectively. In a complete dialogue system, a query is done to a database using the constraints provided by the user when necessary, adding $<$API\_call$>$, $<$DB\_result$>$ and $<$DB\_no\_result$>$ tokens to the messages. 
Given that we do not work with a complete dialogue system, we cannot do the queries, so we added these tokens manually. 
All the messages exchanged in the last nine turns of the conversation were concatenated and the $<$DA\_pred$>$ token is added at the end to indicate to the dialogue policy that a decision must be made.
The source embeddings sequence has the same length as the text dialogue history embeddings sequence, formed by the corresponding $<$user$>$ and $<$system$>$ embeddings, to indicate the source of the token at each position, and a final $<$DA\_pred$>$ token. 
The positional encoding (PE) embeddings sequence is built of vectors with the position information of each token in the sequence. 

\subsection{Baseline architecture}

\begin{figure}[h]
  \centering
  \includegraphics[width=\linewidth]{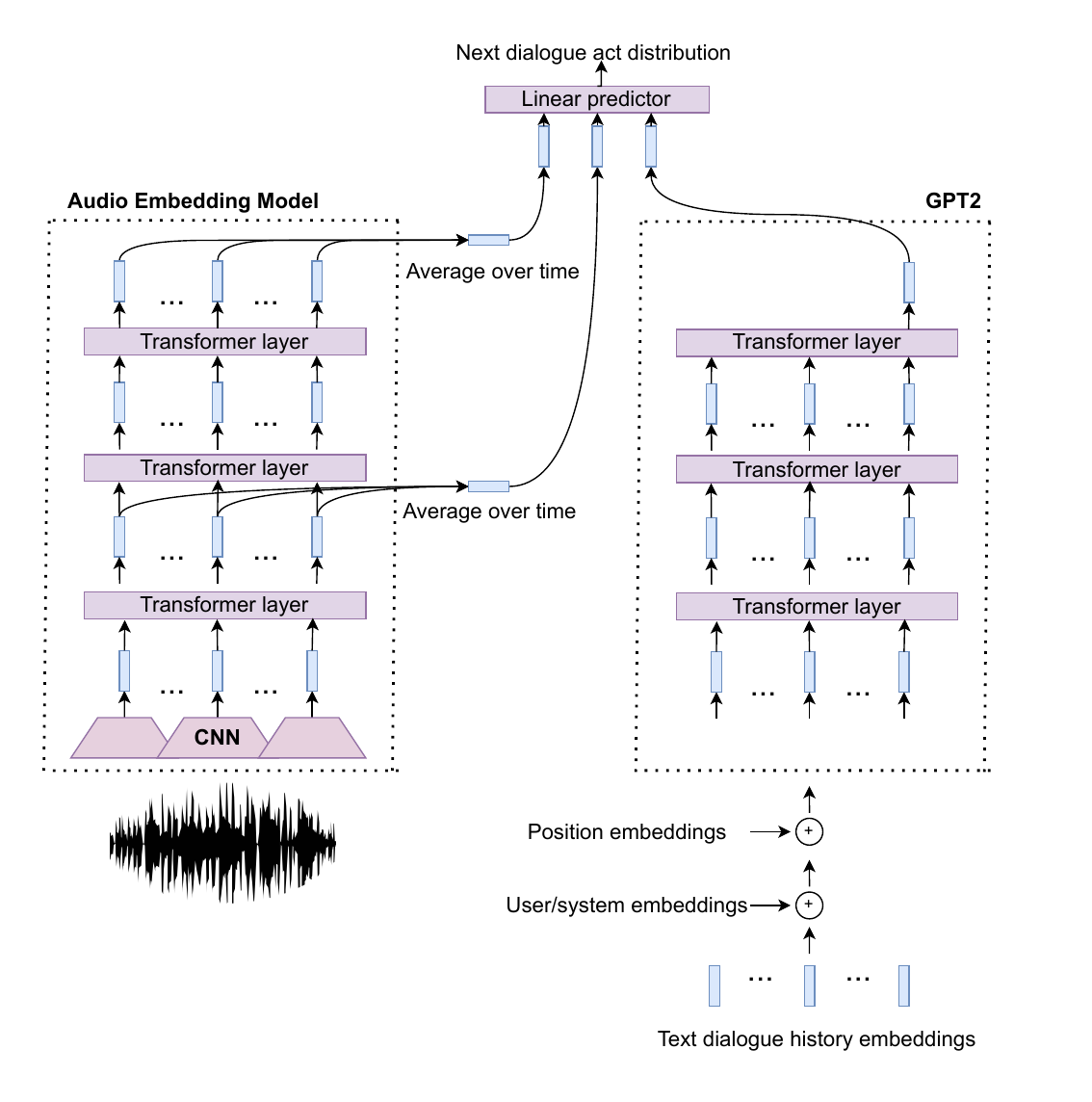}
  \caption{Audio-aware dialogue policy architecture developed in \cite{AEA-2022}.}
  \label{fig:asier_dialogue_policy}
\end{figure}

The dialogue policy is designed as a classifier of a single-label classification problem.
The architecture developed in \cite{AEA-2022}, shown in Figure \ref{fig:asier_dialogue_policy}, will be used as the baseline.
Speech representations are generated by passing last user's turn audio through a pre-trained self-supervised audio model.
A subset of transformer layers is selected, and each of these layers' outputs are averaged over time.
Similarly, text representations of the last nine turns of the conversation are generated using GPT-2.
In this case, only the embedding of the $<$DA\_pred$>$ token generated by GPT-2 from the last transformer layer is selected.
To get the probability distribution of the next dialogue act, the selected text and audio embeddings are concatenated and used as input for a linear predictor. 
The linear predictor consists of a linear layer followed by a softmax layer.
Therefore, the output is a vector containing the probability of each label for the given input.
In \cite{AEA-2022} experiments, they provide evidence that this architecture obtains better performance than using only text (the same architecture, without the audio components).

\subsection{Enhanced audio-aware dialogue policy with Double Multi-Head Attention Multimodal Fusion}

\begin{figure}[h]
  \centering
  \includegraphics[width=\linewidth]{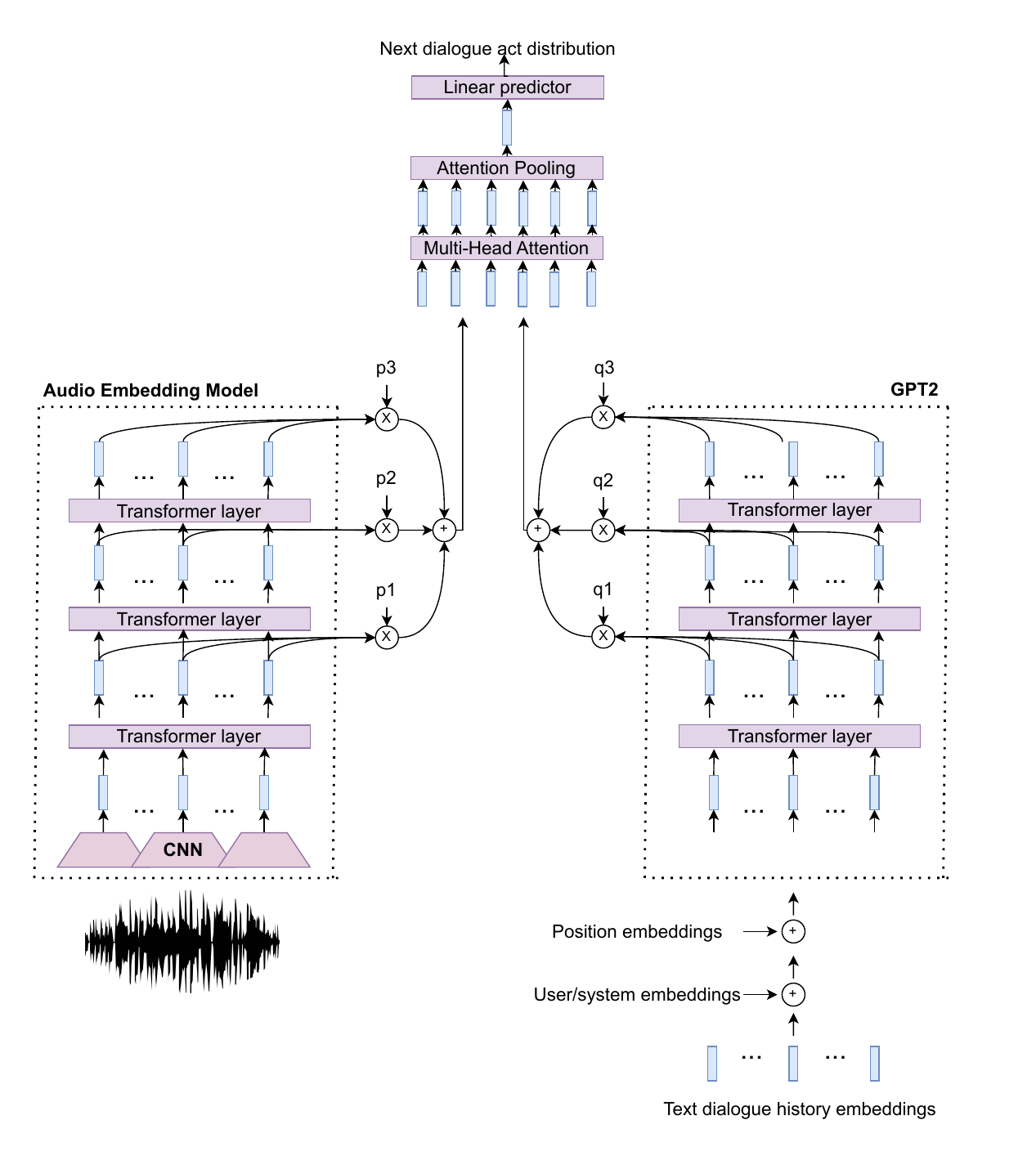}
  \caption{Enhanced audio-aware dialogue policy with Double Multi-Head Attention Multimodal Fusion.}
  \label{fig:double_mha_dialogue_policy}
\end{figure}

In our proposed architecture (Figure \ref{fig:double_mha_dialogue_policy}), speech and text representations are calculated and then combined using a Double Multi-Head Attention component, as in \cite{DMH-2024}.
Speech representations are generated by passing last user's turn audio through a pre-trained self-supervised audio model.
Instead of selecting a subset of transformer layers, every transformer layer is considered by performing a weighted sum.
Also, the sum is done token-wise, so the result is a sequence of vectors with the same length as the number of vectors that each transformer outputs.
This allows the model to learn how much weight to put on each layer, optimizing the architecture to the specific task.
The weights of each layer are learned during network training.
The same procedure is applied to obtain the text representations.
Instead of using only the embedding of the $<$DA\_pred$>$ token generated by GPT-2 from the last transformer layer, a weighted sum of the output of each transformer layer of GPT-2 is performed.
These weights are also learned during the training phase.

Text and speech representations are fused using a Double Multi-Head Attention component.
First, these representations are mixed using an MHA layer to let the model learn complementary information using self-attention scores.
If there are $T_1$ speech representations and $T_2$ text representations, the $T = T_1 + T_2$ representations are input to the first MHA layer.
This first layer will output $T$ multimodal contextualized vectors (independently of the number of heads applied in the MHA). 
Then, a second attention layer is applied to generate a single pooled vector.
Finally, this single pooled vector is used as input for the linear predictor, which is used to obtain the probability distribution of the next dialogue act.

\section{Experiments}

\subsection{Database}

The Dialogue State Tracking Challenge (DSTC) \cite{TDS-2016} is a set of challenges created to compare the performance of different dialogue architectures.
In 2014, DSTC2 was released \cite{TSD-2014}. 
This challenge included a dataset related to restaurant searches, consisting of a series of calls of users to a system to find restaurants that match their preferences, where the user must inform the system about the constraints that the restaurant must satisfy. 
The DSTC2 dataset consists of 3,235 dialogues. 
The dialogues are distributed across train, development, and test partitions.  
There are 1,612 dialogues in the train set, 506 in the development set, and 1,117 in the test set. 
In total, there are 22,266 turns across all conversations, with 10,065 in the train set, 3,428 in the development set, and 8,773 in the test set.

\subsection{Experimental Setup}

The labels to be inferred are all the dialogue acts of the system's next turn. To implement the dialogue policy as a single-label classifier, in each system's turn we concatenate the dialogue acts performed by the system in that turn. By doing so, we end up with 60 labels.
The messages with the $welcomemsg$ label have been removed. 
This is because, in the DSTC2 dataset, the first interaction is always from the system, and it always corresponds to a $welcomemsg$ dialogue act. 
Since inferring that the first dialogue act must be $welcomemsg$ is trivial, we discard it to avoid being overoptimistic when studying the model's performance.

Not all the labels appear in the three dataset partitions.  
The samples from the development and test sets whose labels are not found in the training partition have not been removed, and therefore, they are taken into account during the evaluation step. 
While this might be a serious problem when training a traditional classifier, it might not be as serious when training a dialogue policy. 
This is because, in a dialogue policy, there is usually no single correct response from the system but rather several possible ones. 
To find evidence that making dialogue policies audio-embedding aware can improve performance, we evaluate our models using the User Request Score (URS).
URS assesses the dialogue policy's responsiveness to the user's needs. 
It is defined as the ratio of user requests that the system answers in the message that follows the request.

The DSTC2 dataset provides the correct transcript of the users' audio. 
However, in a real dialogue system, an ASR would be used to transcribe the text. 
To evaluate the dialogue policy with more realistic data, instead of using the database's correct transcriptions of the audios we use the ones provided by an ASR based on a Wav2Vec2 \cite{WAF-2020} network, where the wav2vec2-base-960h checkpoint was fine-tuned using 960 hours of Librispeech \cite{LAA-2015}.
To ensure that all inputs in the batch have the same length, we represent each chat history by a sequence of tokens of the same length. 
This is accomplished using zero padding, with the final token always being $<$DA\_pred$>$. 
Since the GPT-2 architecture allows masking tokens, we mask the padding tokens so that they are not taken into account during forward propagation.

For each turn, the DSTC2 dataset includes the original recorded audio from the user.
We decided to make all audios of 3 seconds length. 
For those larger than 3 seconds, which are $10\%$ of the audios, they have been truncated.
For those shorter than 3 seconds, we applied repetition padding, which is a useful approach, specially in the presence of very short audios \cite{DMH-2024, BAE-2024}. 
Before feeding an audio into an audio embedding model, they have been normalized according to the requirements of each of the models.

The datasets generated during the processing of the DSTC2 dataset were stored in Hugging Face. 
The GPT-2 model, as well as the audio embedding models and the ASR model, correspond to open-source versions available on Hugging Face.
All architectures used in our experiments make use of the Hugging Face implementations of the transformer-based models \cite{HTS-2019}.
The smallest version of GPT-2 is used for text-feature extraction, which has 124 million parameters.
For audio-feature extraction, we experimented with Wav2Vec2.0 \cite{WAF-2020}, HuBERT \cite{HSS-2021}, UniSpeechSAT \cite{UUS-2022} and WavLM \cite{WLS-2022}. 
Specifically, the wav2vec2-base-960h, hubert-base-ls960, unispeech-sat-base and wavlm-base versions, which have approximately 100 million parameters.
Neither GPT-2 nor the models used to generate the audio embeddings are trained from scratch. 
Instead, we used pre-trained models.
On one hand, GPT-2 is pre-trained on WebText, an English corpus of internet websites \cite{LMA-2019}. 
On the other hand, all the audio embedding models are pre-trained on Librispeech, a dataset of read audiobooks in English \cite{LAA-2015}. 
Using audio embedding models pre-trained on the same dataset makes it easier to compare which audio embedding model works better in our task. Additionally, both text and audio datasets are in English, which is the same language as the DSTC2 dialogues.

The first architecture considered in our experiments (Architecture 1) is the only-text one, as in \cite{AEA-2022}.
This is the same as shown in Figure \ref{fig:asier_dialogue_policy}, but only using the text components.
GPT-2 is fine-tuned during training. 
Architecture 2 (Figure \ref{fig:asier_dialogue_policy}) is the audio-aware dialogue policy as in \cite{AEA-2022}.
The GPT-2 model used corresponds to the one obtained after training Architecture 1, and it is frozen. 
In \cite{AEA-2022} experiments, Wav2Vec2.0, HuBERT or UniSpeechSAT were used as audio embedding models.
Additionally, we experimented with WavLM, which has been proven to be an effective speech representations extractor \cite{WLS-2022}.
The selected audio embedding model is frozen.
In the WavLM case, we conducted experiments to determine which transformer layer (or combination of them) outputs the best audio embedding to be used as input for the linear predictor. 
Architecture 1 and 2 will be used as baselines for our experiments.

For ablation purposes, we considered experimenting with Architecture 3, which is the same as in Figure \ref{fig:double_mha_dialogue_policy}, but only using the text components.
GPT-2 is fine-tuned during training. 
Instead of using only the embedding of the $<$DA\_pred$>$ token generated by GPT-2 from the last transformer layer, a weighted sum of the output of each transformer layer of GPT-2 is performed, with weights learned during the training phase.
The MHA layer in the Double Multi-Head Attention component uses 4 heads.

Our proposed architecture (Architecture 4) is shown in Figure \ref{fig:double_mha_dialogue_policy}.
The GPT-2 model is fine-tuned during training.
We experimented using four different frozen pre-trained self-supervised audio embedding models: Wav2Vec2.0, HuBERT, UniSpeechSAT or WavLM. 
Both for text and audio, instead of selecting a subset of transformer layers, every transformer layer is considered by performing a weighted sum.
Also, the sum is done token-wise, so the result is a sequence of vectors with the same length as the number of vectors that each transformer outputs.
The weights of each layer are learned during network training.
The MHA layer in the Double Multi-Head Attention component uses 4 heads.

For statistical purposes, each experiment is conducted several times, each time initializing randomly the parameters to be learned.
Architecture 1 is trained 3 times, leading to 3 fine-tuned GPT-2. 
For each fine-tuned GPT-2, Architecture 2 is trained 3 times for each audio embedding model.  
Similarly, Architecture 3 is trained 3 times, leading to 3 fine-tuned GPT-2. 
For each fine-tuned GPT-2, Architecture 4 is trained 3 times for each audio embedding model.  

To make the experiments as comparable as possible to those in \cite{AEA-2022}, the loss function used is the cross-entropy loss. 
Given that the label distribution is unbalanced (80\% of the samples correspond to the top 15 most common labels) some experiments with the weighted cross-entropy loss were also conducted.
Nevertheless, the results obtained across all audio embedding models were worse than using the unweighted cross-entropy loss.

To conduct the experiments, we utilized Google Colab \cite{GOC-2024}, a hosted Jupyter Notebook platform designed for machine learning and data science tasks.
Throughout all the experiments, we made use of an NVIDIA T4 GPU.

\subsection{Results}

The results obtained using architectures 1 to 4 are shown in Table \ref{tab:results_asr_transcript}.
Architecture 2 obtains, on average, a $2.7\%$ URS relative improvement with respect to architecture 1.
Both architectures 1 and 2 were developed in \cite{AEA-2022}, and these results verifies the fact that adding audio information to the dialogue policy improves the performance, as stated in their work.
In our experiments, the best audio model for architecture 2 is HuBERT, achieving a $3.5\%$ URS relative improvement with respect to architecture 1.

Our proposed architecture (Architecture 4) outperforms text-only architectures obtaining, on average, a $6.8\%$ URS relative improvement with respect to architecture 1.
This shows that adding audio information to the dialogue policy can be crucial to improving performance in these systems.
Given that architecture 4 obtains a $6.2\%$ URS relative improvement with respect to architecture 3, we can conclude that this improvement is not driven by the modifications added to the text feature extractor.
In our experiments, the best audio model for Architecture 4 is UniSpeechSAT, achieving a $9.8\%$ URS relative improvement with respect to architecture 1, and a $9.2\%$ URS relative improvement with respect to architecture 3.
Also, our proposed architecture outperforms the audio-aware dialogue policy architecture developed in \cite{AEA-2022} obtaining, on average, a $3.1\%$ URS relative improvement with respect to Architecture 2 with HuBERT.
Specifically, Architecture 4 with UniSpeechSAT obtaining a $6.0\%$ URS relative improvement with respect to Architecture 2 with HuBERT. 

These results show that our Double Multi-Head Attention Multimodal Fusion component enhance the audio-aware dialogue policy performance, allowing the model to capture complex relationships between speech and text representations.
These relationships are computed by jointly attending to information from different representations using attention scores.
Using a Multi-Head Attention layer allows to create several contextual representations combining information in different ways.

\begin{table}[h!]
\vspace{2mm}
    \centering
    \renewcommand{\arraystretch}{1.2}
    \begin{tabular}{|l|l|c|}
        \hline
        \textbf{Architecture} & \textbf{Audio Embedding} & \textbf{URS} \\ 
        \hline
        Architecture 1 \cite{AEA-2022} & Text-only & $77.575  \pm 0.266$ \\ 
        \hline
        \multirow[c]{4}{*}{Architecture 2 \cite{AEA-2022}} & Wav2Vec2.0 & $80.044 \pm 0.646$ \\
         & HuBERT & $80.318 \pm 0.931$ \\
         & UniSpeechSAT & $79.744 \pm 0.635$ \\
        & WavLM & $78.471 \pm 0.420$ \\
        \hline
        Architecture 3 & Text-only & $77.972 \pm 1.841$ \\
        \hline
        \multirow[c]{4}{*}{Architecture 4} & Wav2Vec2.0 & $81.789 \pm 2.280$ \\
         & HuBERT & $81.856 \pm 1.881$ \\
         & UniSpeechSAT & $85.156 \pm 1.799$ \\
         & WavLM & $82.452 \pm 0.915$ \\
        \hline
    \end{tabular}
    \vspace{2mm}
    \caption{Average and standard deviation of the User Request Score (URS) obtained for each architecture.}
    \label{tab:results_asr_transcript}
\end{table}

\section{Conclusions}

In this paper, we have proposed an architecture that uses audio to improve dialogue policy performance.
Speech and text representations were extracted using pre-trained self-supervised models.
These multimodal representations are mixed adopting an early fusion strategy by using a Double Multi-Head Attention Multimodal Fusion component.
First, a Multi-Head Attention layer generates complementary contextualized representations. 
A second attention layer is then applied to pool these representations into an utterance-level vector. 
This component allows the model to capture complex relationships jointly attending to information from different representation subspaces. 
Our proposed architecture outperforms text-only architectures achieving a $9.8\%$ User Request Score relative improvement. 

\section{Acknowledgements}

This work has been promoted and financed by the Generalitat de Catalunya through the Aina project and by the Spanish Ministerio de Ciencia e Innovación through the project AdaVoice PID2019-107579RB-I00.
The second author is supported by a FI grant from the Catalan government.

\bibliographystyle{IEEEtran}
\bibliography{mybib}

\begin{thebibliography}{10}
\providecommand{\url}[1]{#1}
\csname url@samestyle\endcsname
\providecommand{\newblock}{\relax}
\providecommand{\bibinfo}[2]{#2}
\providecommand{\BIBentrySTDinterwordspacing}{\spaceskip=0pt\relax}
\providecommand{\BIBentryALTinterwordstretchfactor}{4}
\providecommand{\BIBentryALTinterwordspacing}{\spaceskip=\fontdimen2\font plus
\BIBentryALTinterwordstretchfactor\fontdimen3\font minus \fontdimen4\font\relax}
\providecommand{\BIBforeignlanguage}[2]{{%
\expandafter\ifx\csname l@#1\endcsname\relax
\typeout{** WARNING: IEEEtran.bst: No hyphenation pattern has been}%
\typeout{** loaded for the language `#1'. Using the pattern for}%
\typeout{** the default language instead.}%
\else
\language=\csname l@#1\endcsname
\fi
#2}}
\providecommand{\BIBdecl}{\relax}
\BIBdecl

\bibitem{DSA-2013}
S.~Arora, K.~Batra, and S.~Singh, ``Dialogue system: A brief review,'' \emph{arXiv preprint arXiv:1306.4134}, 2013.

\bibitem{VID-2017}
A.~Das, S.~Kottur, K.~Gupta, A.~Singh, D.~Yadav, J.~M. Moura, D.~Parikh, and D.~Batra, ``Visual dialog,'' in \emph{Proceedings of the IEEE conference on computer vision and pattern recognition}, 2017, pp. 326--335.

\bibitem{NAT-2018}
J.~Gao, M.~Galley, and L.~Li, ``Neural approaches to conversational ai,'' in \emph{The 41st international ACM SIGIR conference on research \& development in information retrieval}, 2018, pp. 1371--1374.

\bibitem{CBG-2024}
Z.~Liu, W.~Ping, R.~Roy, P.~Xu, M.~Shoeybi, and B.~Catanzaro, ``Chatqa: Building gpt-4 level conversational qa models,'' \emph{arXiv preprint arXiv:2401.10225}, 2024.

\bibitem{ASO-2017}
H.~Chen, X.~Liu, D.~Yin, and J.~Tang, ``A survey on dialogue systems: Recent advances and new frontiers,'' \emph{Acm Sigkdd Explorations Newsletter}, vol.~19, no.~2, pp. 25--35, 2017.

\bibitem{RAI-2023}
J.~Ni, T.~Young, V.~Pandelea, F.~Xue, and E.~Cambria, ``Recent advances in deep learning based dialogue systems: A systematic survey,'' \emph{Artificial intelligence review}, vol.~56, no.~4, pp. 3055--3155, 2023.

\bibitem{OTO-2019}
C.~Tao, W.~Wu, C.~Xu, W.~Hu, D.~Zhao, and R.~Yan, ``One time of interaction may not be enough: Go deep with an interaction-over-interaction network for response selection in dialogues,'' in \emph{Proceedings of the 57th annual meeting of the association for computational linguistics}, 2019, pp. 1--11.

\bibitem{SCF-2020}
S.~Coope, T.~Farghly, D.~Gerz, I.~Vuli{\'c}, and M.~Henderson, ``Span-convert: Few-shot span extraction for dialog with pretrained conversational representations,'' \emph{arXiv preprint arXiv:2005.08866}, 2020.

\bibitem{ANB-2016}
T.-H. Wen, D.~Vandyke, N.~Mrksic, M.~Gasic, L.~M. Rojas-Barahona, P.-H. Su, S.~Ultes, and S.~Young, ``A network-based end-to-end trainable task-oriented dialogue system,'' \emph{arXiv preprint arXiv:1604.04562}, 2016.

\bibitem{ETE-2020}
D.~Ham, J.-G. Lee, Y.~Jang, and K.-E. Kim, ``End-to-end neural pipeline for goal-oriented dialogue systems using gpt-2,'' in \emph{Proceedings of the 58th annual meeting of the association for computational linguistics}, 2020, pp. 583--592.

\bibitem{LMA-2019}
A.~Radford, J.~Wu, R.~Child, D.~Luan, D.~Amodei, I.~Sutskever \emph{et~al.}, ``Language models are unsupervised multitask learners,'' \emph{OpenAI blog}, vol.~1, no.~8, p.~9, 2019.

\bibitem{EEO-2013}
T.~Mikolov, K.~Chen, G.~Corrado, and J.~Dean, ``Efficient estimation of word representations in vector space,'' \emph{arXiv preprint arXiv:1301.3781}, 2013.

\bibitem{GGV-2014}
J.~Pennington, R.~Socher, and C.~D. Manning, ``Glove: Global vectors for word representation,'' in \emph{Proceedings of the 2014 conference on empirical methods in natural language processing (EMNLP)}, 2014, pp. 1532--1543.

\bibitem{HCA-2019}
K.~Ethayarajh, ``How contextual are contextualized word representations? comparing the geometry of bert, elmo, and gpt-2 embeddings,'' \emph{arXiv preprint arXiv:1909.00512}, 2019.

\bibitem{LKA-2019}
N.~F. Liu, M.~Gardner, Y.~Belinkov, M.~E. Peters, and N.~A. Smith, ``Linguistic knowledge and transferability of contextual representations,'' \emph{arXiv preprint arXiv:1903.08855}, 2019.

\bibitem{DCW-2018}
M.~E. Peters, M.~Neumann, M.~Iyyer, M.~Gardner, C.~Clark, K.~Lee, and L.~Zettlemoyer, ``Deep contextualized word representations,'' in \emph{Proceedings of the 2018 Conference of the North {A}merican Chapter of the Association for Computational Linguistics: Human Language Technologies, Volume 1 (Long Papers)}, 2018, pp. 2227--2237.

\bibitem{BPT-2018}
J.~Devlin, M.-W. Chang, K.~Lee, and K.~Toutanova, ``Bert: Pre-training of deep bidirectional transformers for language understanding,'' \emph{arXiv preprint arXiv:1810.04805}, 2018.

\bibitem{AEH-2021}
A.~L. Zorrilla, M.~I. Torres, and H.~Cuay{\'a}huitl, ``Audio embeddings help to learn better dialogue policies,'' in \emph{2021 IEEE Automatic Speech Recognition and Understanding Workshop (ASRU)}.\hskip 1em plus 0.5em minus 0.4em\relax IEEE, 2021, pp. 962--968.

\bibitem{AEA-2022}
{Zorrilla, Asier L{\'o}pez and Torres, Mar{\'\i}a In{\'e}s and Cuay{\'a}huitl, Heriberto}, ``Audio embedding-aware dialogue policy learning,'' \emph{IEEE/ACM Transactions on Audio, Speech, and Language Processing}, vol.~31, pp. 525--538, 2022.

\bibitem{DSW-2020}
T.~Young, V.~Pandelea, S.~Poria, and E.~Cambria, ``Dialogue systems with audio context,'' \emph{Neurocomputing}, vol. 388, pp. 102--109, 2020.

\bibitem{ETF-2023}
M.~Paw{\l}owski, A.~Wr{\'o}blewska, and S.~Sysko-Roma{\'n}czuk, ``Effective techniques for multimodal data fusion: A comparative analysis,'' \emph{Sensors}, vol.~23, no.~5, p. 2381, 2023.

\bibitem{DMH-2021}
M.~India, P.~Safari, and J.~Hernando, ``Double multi-head attention for speaker verification,'' in \emph{ICASSP 2021-2021 IEEE International Conference on Acoustics, Speech and Signal Processing (ICASSP)}.\hskip 1em plus 0.5em minus 0.4em\relax IEEE, 2021, pp. 6144--6148.

\bibitem{DMH-2024}
F.~Costa, M.~India, and J.~Hernando, ``{Double Multi-Head Attention Multimodal System for Odyssey 2024 Speech Emotion Recognition Challenge},'' in \emph{Proc. The Speaker and Language Recognition Workshop (Odyssey 2024)}, 2024, pp. 266--273.

\bibitem{BAE-2024}
M.~Casals-Salvador, F.~Costa, M.~India, and J.~Hernando, ``{BSC-UPC at EmoSPeech-IberLEF2024: Attention Pooling for Emotion Recognition},'' \emph{arXiv preprint arXiv:2407.12467}, 2024.

\bibitem{WAF-2020}
A.~Baevski, Y.~Zhou, A.~Mohamed, and M.~Auli, ``wav2vec 2.0: A framework for self-supervised learning of speech representations,'' \emph{Advances in neural information processing systems}, vol.~33, pp. 12\,449--12\,460, 2020.

\bibitem{HSS-2021}
W.-N. Hsu, B.~Bolte, Y.-H.~H. Tsai, K.~Lakhotia, R.~Salakhutdinov, and A.~Mohamed, ``Hubert: Self-supervised speech representation learning by masked prediction of hidden units,'' \emph{IEEE/ACM transactions on audio, speech, and language processing}, vol.~29, pp. 3451--3460, 2021.

\bibitem{UUS-2022}
S.~Chen, Y.~Wu, C.~Wang, Z.~Chen, Z.~Chen, S.~Liu, J.~Wu, Y.~Qian, F.~Wei, J.~Li \emph{et~al.}, ``Unispeech-sat: Universal speech representation learning with speaker aware pre-training,'' in \emph{ICASSP 2022-2022 IEEE International Conference on Acoustics, Speech and Signal Processing (ICASSP)}.\hskip 1em plus 0.5em minus 0.4em\relax IEEE, 2022, pp. 6152--6156.

\bibitem{WLS-2022}
S.~Chen, C.~Wang, Z.~Chen, Y.~Wu, S.~Liu, Z.~Chen, J.~Li, N.~Kanda, T.~Yoshioka, X.~Xiao \emph{et~al.}, ``Wavlm: Large-scale self-supervised pre-training for full stack speech processing,'' \emph{IEEE Journal of Selected Topics in Signal Processing}, vol.~16, no.~6, pp. 1505--1518, 2022.

\bibitem{TDS-2016}
J.~D. Williams, A.~Raux, and M.~Henderson, ``The dialog state tracking challenge series: A review,'' \emph{Dialogue \& Discourse}, vol.~7, no.~3, pp. 4--33, 2016.

\bibitem{TSD-2014}
M.~Henderson, B.~Thomson, and J.~D. Williams, ``The second dialog state tracking challenge,'' in \emph{Proceedings of the 15th annual meeting of the special interest group on discourse and dialogue (SIGDIAL)}, 2014, pp. 263--272.

\bibitem{LAA-2015}
V.~Panayotov, G.~Chen, D.~Povey, and S.~Khudanpur, ``Librispeech: an asr corpus based on public domain audio books,'' in \emph{2015 IEEE international conference on acoustics, speech and signal processing (ICASSP)}.\hskip 1em plus 0.5em minus 0.4em\relax IEEE, 2015, pp. 5206--5210.

\bibitem{HTS-2019}
T.~Wolf, L.~Debut, V.~Sanh, J.~Chaumond, C.~Delangue, A.~Moi, P.~Cistac, T.~Rault, R.~Louf, M.~Funtowicz \emph{et~al.}, ``Huggingface's transformers: State-of-the-art natural language processing,'' \emph{arXiv preprint arXiv:1910.03771}, 2019.

\bibitem{GOC-2024}
Google, ``Google colaboratory,'' 2024, \url{https://colab.research.google.com/} [Retrieved: July 24, 2024].

\end{thebibliography}

\end{document}